# Emission from quantum-dot high-β microcavities: transition from spontaneous emission to lasing and the effects of superradiant emitter coupling

S. Kreinberg,[1] W. W. Chow,[2] J. Wolters,[1] C. Schneider,[3] C. Gies,[4] F. Jahnke,[4] S. Höfling,[3,4] M. Kamp[3] and S. Reitzenstein[1]

[1] *Institut für Festkörperphysik, Technische Universität Berlin, 10623 Berlin, Germany*
[2] *Sandia National Laboratories, Albuquerque, NM 87185-1086, U.S.A.*
[3] *Lehrstuhl für Technische Physik, Universität Würzburg, 97074 Würzburg, Germany*
[4] *School of Physics and Astronomy, University of St Andrews, St Andrews, KY16 9SS, UK*
[5] *Institute for Theoretical Physics, University of Bremen, 28334 Bremen, Germany*

Corresponding author:
W. W. Chow, phone: (505)844-9088, fax: (505)844-3211, Email: wwchow@sandia.gov

**ABSTRACT:** Measured and calculated results are presented on the emission properties of a new class of emitters operating in the cavity quantum electrodynamics regime. The structures are based on high-finesse GaAs/AlAs micropillar cavities, each with an active medium consisting of a layer of InGaAs quantum dots and distinguishing feature of having substantial fraction of spontaneous emission channeled into one cavity mode (high β-factor). This paper shows that the usual criterion for lasing with a conventional (low β-factor) cavity, a sharp non-linearity in an input-output curve accompanied by noticeable linewidth narrowing, has to be reinforced by the equal-time second-order photon autocorrelation function for confirming lasing. It will also show that the equal-time second-order photon autocorrelation function is useful for recognizing superradiance, a manifestation of the correlations possible in high-β microcavities operating with quantum dots. In terms of consolidating the collected data and identifying the physics underlying laser action, both theory and experiment suggest a sole dependence on intracavity photon number. Evidence for this comes from all our measured and calculated data on emission coherence and fluctuation, for devices ranging from LEDs and cavity-enhanced LEDs to lasers, lying on the same two curves: one for linewidth narrowing versus intracavity photon number and the other for $g^{(2)}(0)$ versus intracavity photon number.

# INTRODUCTION

The question, 'What is a laser?' has been a long-standing subject of discussion.[1] For example, during the 1970's, there was much debate on whether the x-ray laser, if invented, was really a 'laser' given the poor or non-existent resonator. Now, the question is resurfacing, but for an entirely opposite reason. With micro- and nano-cavities, giving high-Q factors and close to complete channeling of all spontaneous emission into a single cavity mode ($\beta \lesssim 1$), cavity quantum electrodynamics (CQED) effects may be present to the extent that the concept of lasing action and its verification have to be reexamined.[2-5]

The development of low-threshold micro- and nano-lasers has become an important interdisciplinary research topic in recent years, compassing expitaxy growth and lithographic chemistry, spectroscopic and photon correlation measurements, as well as quantum optics and quantum electronics physics.[6,7] The research is conducted in multiple material systems, ranging from III-V and III-nitride compounds to silicon.[8,9] Even emerging 2D materials have been applied as active medium of nanolasers. Cavity designs for photonic lattices, micropillars and plasmonic cavities embedding bulk, quantum-well and quantum-dot gain material are being explored. [10,11]

There is considerable activity focusing on high-β lasers, because of the potential for exceeding present limits on lasing threshold, modulation speeds and spatial footprint. Advances are being made in geometrical waveguiding, and in enhancing their performance by quantum optical effects such as the Purcell effect.[12] Interests in using high-β cavities extend beyond lasers to light-emitting diodes (LEDs) for applications such as high-efficiency lighting and to single-photon sources for quantum information processing.[13,14] In research, high-β emitters provide experimental platforms to study the intricate interplay between classical cavity-mode confinement and quantum optics.[15-17]

Along with the exciting progress achieved in device performance and in realizing application potential, there is a vivid debate on the precise definition and verification of lasing in high-$\beta$ emitters.[18-21] This is especially true in the limiting regime involving the interesting prospect of thresholdless lasing.[22-25] Indeed, with increasing $\beta$-factor it becomes more difficult to find a definitive transition in device output from being spontaneous emission dominated to being governed by stimulated emission. Such a transition has long been accepted as the characteristic signature of lasing.

Usually, lasing in a high-$\beta$ emitter is claimed when there is noticeable linewidth narrowing (or coherence time increase) accompanied by an output versus input curve (often with output in arbitrary units) that does not show a pronounced 'S' shape in a log-log plot. In this paper, we will show that these two pieces of information are insufficient proof for lasing. Our study involves measuring and modeling emitters consisting of AlAs/GaAs micro-pillar cavities with GaInAs quantum-dot (QD) gain regions. The device configurations vary sufficiently to cover emitters operating purely as LEDs, as lasers and as cavity-enhanced LEDs. Many of the devices show input-output curves and linewidth narrowing that satisfy the generally accepted criterion for lasing in high-$\beta$ cavities. However, measurements of second-order photon autocorrelation indicate distinctly different statistical properties of photons in the emissions.

Photon correlations have been largely ignored by the laser engineering community when it comes to characterizing devices. Besides being time and equipment demanding, the reason is that the measurement is unnecessary for emitters operating with conventional resonators. There, the customary lasing criterion is the appearance of a noticeable kink in a log-log plot of output versus input power. For further confirmation, one looks for a sharp increase in linewidth narrowing in the vicinity of the input-output jump. However, the abruptness of these signatures diminish in high-β lasers, with input-output jump disappearing all together and the linewidth narrowing not approaching the Schawlow-Townes linewidth in the limit of 'thresholdless' lasing.[26-29] This makes for much discussion when distinguishing between high-β lasing and effects from nonlinearities (e.g. from saturation or state filling) in a non-lasing, $\beta \ll 1$ emitter.

In contrast to laser engineering, the quantum optics community has long studied photon statistics, which is customarily gauged by the equal-time second-order photon auto-correlation function $g^{(2)}(0)$.[30-32] Specifically, when $g^{(2)}(0)$ reduces from 2 to 1, the emitted light transitions from spontaneous emission dominated to predominately stimulated emission. This change from thermal to coherent emission serves as a reliable indicator of lasing.

Section II in this paper describes the experimental devices and setup for performing systematic measurements of emission intensity, coherence and photon correlations. An overview of the theoretical approach to modeling the quantum optics of nano-emitters is given in Sec. III. In Sec. IV, we present results on output intensity, emission linewidth and $g^{(2)}(0)$ versus optical pump power for micro-pillar emitters with Q-factors varying from 8000 to 35000, β-factors from 0.2 to 0.4 and quantum-dot numbers from 5 to 40. Analyses of the experiments, using the CQED model, show a commonality among the devices, in the form of a relationship between $g^{(2)}(0)$ and the intracavity photon number in the lasing mode. The origin of this link, which comes directly from the laser-field quantization, will be discussed, together with a quantitative connection between linewidth narrowing and $g^{(2)}(0)$.

## MATERIALS AND METHODS

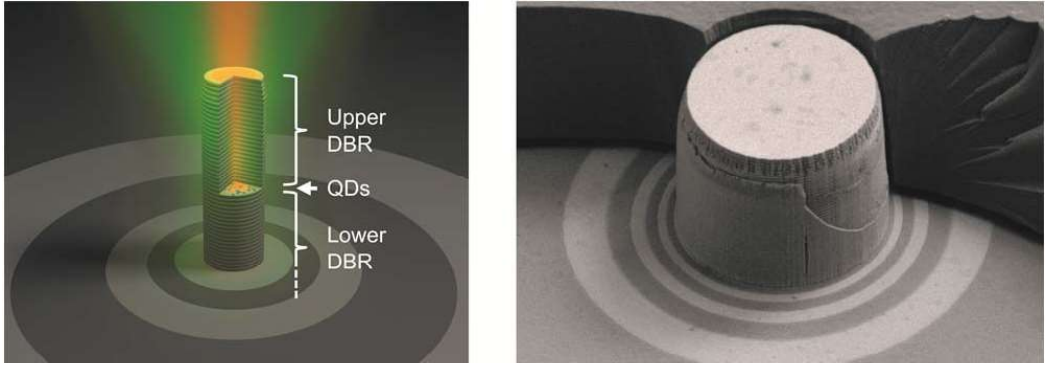

**Figure 1.** **(Left)** Artist's impression of an optically pumped quantum-dot micropillar consisting of two GaAs/AlAs distributed Bragg reflectors (DBRs) and quantum dots (QDs) on a wetting layer. **(Right)** Scanning electron microscopy (SEM) image of 8 μm diameter micropillar, where part of the planarizing BCB layer (visible in the background) was mechanically removed to show the layers comprising the micropillar's DBRs.

This section describes the emitter fabrication and measurement setup used for studying emission from low-mode-volume, high-β micropillars. Employing electron-beam lithography and plasma reactive ion etching, each free-standing micropillar is processed from the same planar high-Q microcavity consisting of a lower and an upper AlAs/GaAs diffractive Bragg reflector (DBR) with 25 and 30 layer pairs, respectively, and a single layer of InGaAs QDs as active medium (see Fig. 1). The structures were planarized by the polymer BCB for mechanical stability and to suppress oxidation of the AlAs layers in the DBRs. Owing to a slight radial asymmetry in heterostructure layer thicknesses, different locations on the epitaxial wafer give micropillars with different detuning between the fundamental cavity mode and QD resonance, thus allowing control of modal gain. In addition, with varying micropillar diameters, we can influence the Purcell enhancement of spontaneous emission, and thus the β-factor via its dependence on modal volume and cavity Q factor. In this way, we are able to fabricate emitters showing only luminescence (LED behavior), cavity-enhanced spontaneous emission, and high-β lasing. Details on the micropillar layout and processing have been reported earlier.[33]

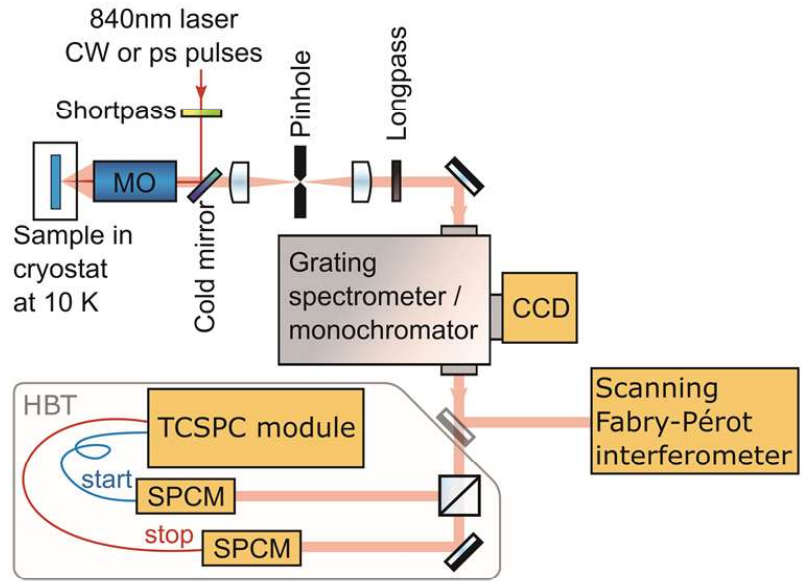

**Figure 2.** Experimental setup. At the top left corner is a cryostat containing a sample that is pumped by a titanium:sapphire laser. The steady-state, constant wave (cw) pumped photoluminescence is analyzed with a grating spectrometer connected to a charge-coupled device (CCD), a scanning Fabry-Perot interferometer, a Hanbury Brown and Twiss (HBT) setup with single-photon counting modules (SPCMs) and a time-correlated single-photon counting (TCSPC) module. The cold mirror superimposes the pump laser onto the detection path, the pinhole provides spatial filtering, and the 850nm longpass filter blocks scattered pump laser light.

Figure 2 is a schematic view of the spectroscopic measurement setup. Experiments are performed at 10 K by mounting the micro-pillar sample on a cold-finger of a liquid-helium flow cryostat. Continuous wave (cw) excitation is provided by an external-cavity laser tuned to 840nm, which resonances with the wetting layer of QDs in the active region. µPL emission spectra are measured either by a grating spectrometer with a spectral resolution of 25 µeV, or by a scanning Fabry-Pérot interferometer with a free-spectral range of 30 µeV and spectral resolution of 0.5 μeV. The second-order photon auto-correlation function is measured with a fiber-coupled Hanbury-Brown and Twiss (HBT) configuration equipped with Si-avalanche photodiode (APD) based single-photon counting modules (SPCMs) with a temporal resolution of 260ps.

A CQED model is used to compute the intracavity photon number, emission linewidth and 2$^{nd}$ order intensity correlation function for different pump powers and QD-micropillar configurations. The model is derived in the Heisenberg Picture using a cluster expansion method to obtain a closed set of equations of motion for the polarization $p_n$ photon population $n_p$, and electron (hole) carrier population $n_{e,n}^{QD}$ ($n_{h,n}^{QD}$):[32,34,35]

$$\frac{dp_n}{dt} = -[i\,(\nu - \omega_n) + (\gamma + \gamma_c)]\,p_n + g\left[n_{e,n}^{QD} n_{h,n}^{QD} + \left(n_{e,n}^{QD} + n_{h,n}^{QD} - 1\right)n_p\right], \quad (1)$$

$$\frac{dn_p}{dt} = 2\sum_n n_{inh}(\omega_n)\,Re(p_n) - 2\gamma_c n_p, \quad (2)$$

$$\frac{dn_{\sigma,n}^{QD}}{dt} = -2Re(p_n) - \gamma_{nl} n_{e,n}^{QD} n_{h,n}^{QD} - \gamma_{nr} n_{\sigma,n}^{QD} + \frac{\eta P}{\hbar\omega N_\sigma^p}\,f\bigl(\varepsilon_{\sigma,n}, \mu_\sigma^p, T_p\bigr)(1 - n_{\sigma,k}^{QW})$$
$$-\gamma_{c-c}\left[n_{\sigma,n}^{QD} - f\bigl(\varepsilon_{\sigma,n}, \mu_\sigma, T\bigr)\right] - \gamma_{c-p}\left[n_{\sigma,n}^{QD} - f\bigl(\varepsilon_{\sigma,n}, \mu_\sigma^l, T_l\bigr)\right]. \quad (3)$$

Input parameters are the cavity photon decay rate $2\gamma_c$, the spontaneous emission rate into nonlasing modes $\gamma_{nl}$ and the nonradiative carrier-loss rate $\gamma_{nr}$. The subscript σ = e (h) labels the electron (hole) population and single-particle energy $\varepsilon_{\sigma,n}$, $g$ is the light-matter coupling coefficient, and the pump laser is described by its power $P$ and photon energy $\hbar\omega$. We assume that the population that eventually enters the QD state is described by a Fermi-Dirac function $f\bigl(\varepsilon_{\sigma,k}^{QW}, \mu_\sigma^p, T_p\bigr)$ with chemical potential $\mu_\sigma^p$, temperature $T_p$, giving a total population of $N_\sigma^p = \sum_k f\bigl(\varepsilon_{\sigma,k}^{QW}, \mu_\sigma^p, T_p\bigr)$. The efficiency in populating the QD state from the wetting layer states is η. In the above equations scattering effects are approximated by the terms containing the effective dephasing, carrier-carrier and carrier-phonon coefficients $\gamma$, $\gamma_{c-c}$ and $\gamma_{c-p}$. Also in those terms are the asymptotic Fermi-Dirac distributions approached through scattering $f(\varepsilon_{\sigma,n}, \mu_\sigma, T)$ and $f(\varepsilon_{\sigma,k}^{QW}, \mu_\sigma, T)$. For the emission linewidth and equal-time 2$^{nd}$ order intensity correlation function, we use $\Delta\omega = \left(2\int_{-\infty}^{\infty} d\tau \left|g^{(1)}(\tau)\right|^2\right)^{-1}$ and $g^{(2)}(0) = \langle b^\dagger b^\dagger bb \rangle / n_p^2$, respectively, where $g^{(1)}(\tau) = \langle b^\dagger b(\tau) \rangle / n_p$, and the correlations $\langle b^\dagger b(\tau) \rangle$ and $\langle b^\dagger b^\dagger bb \rangle$, involving photon annihilation and creation operators $b$ and $b^\dagger$, are evaluated under stationary situation by solving the next level of equations of motion in the cluster expansion. The $\beta$-factor is computed from the rates of emission into the laser mode and into nonlasing modes according to the system parameters.[23] Furthermore, for some emitters, the electron-hole polarization receives contributions from carrier correlations such as $\langle v_n^\dagger c_m^\dagger c_m v_n \rangle$, where $c_m$ ($v_n$) and $c_m^\dagger$ ($v_n^\dagger$) are electron (hole) creation and annihilation operators. These correlations give rise to coherence phenomena such as subradiance and superradiance, which modify the input-output and $g^{(2)}(0)$ behaviors of an emitter, as will be discussed in the next section.[36,37,38]

Table 1. Parameters for micropillars A-E used in the study.

|   | ∅ (μm) | Q | β | $N_{QD}$ |
|---|---|---|---|---|
| A | 1.7 | 8300 | 0.40 | 10 |
| B | 2.0 | 32100 | 0.37 | 6 |
| C | 2.0 | 32100 | 0.37 | 15 |
| D | 2.5 | 22800 | 0.23 | 60 |
| E | 2.5 | 24900 | 0.72 | 40 |

**RESULTS AND DISCUSSION**

In this section, we present results from measuring and modeling of five micropillars A-E. Table I lists the device parameters relevant to emission properties. The Q-factors in column 3 are obtained directly from experiment via dividing the measured emission energy by the spectral linewidth at the low-excitation plateau region. These Q-values are used in the modeling. The β-factor and number of QDs resonant with the lasing-mode resonance $N_{QD}$ are extracted by simultaneously fitting the theoretical and experiment curves for intensity and linewidth versus pump power. We would like to note that for the given areal density of QDs ($2 \times 10^9$ cm$^2$) the micropillar diameter determines the absolute number of QDs in the active layer. The effective number of QDs $N_{QD}$ depends also on the spectral overlap between the resonator mode and the inhomogenously broadened QD emission band, which explains e.g. that $N_{QD}$ is higher for emitter A if compared to emitter B with larger diameter. Two of the devices (micropillars A and B, with 1.7μm and 2.0 μm diameters, respectively) operate purely as LEDs; one exhibits behaviors of a cavity-enhanced LED (micropillar C, with 2.0 μm diameter); the remaining two are lasers (micropillars D and E, with 2.5μm diameters), with micropillar E exhibiting subradiance and then superradiance emission during the transition from spontaneous emission to lasing. The assignments are based on the combination of emission properties shown in Figs. 2, 3 and 5, which we will discuss in the remainder of this section.

First, we look at the steady-state input-output relationship, which is an expedient and therefore frequently employed method for demonstrating lasing in conventional ($\beta \ll 1$) lasers. In Fig. 3, the data points are from experiment and the grey curves are computed using the CQED model. The matching of experimental and theoretical curves allows the extraction of the spontaneous emission rate into nonlasing modes $\gamma_{nl}$, QD number $N_{QD}$, and carrier injection efficiency η for each micropillar. The comparison also provides a calibration of the detector setup, allowing the conversion from detector counts over the integration period to intracavity photon number in the lasing mode (left and right ordinates in top row of Fig. 3). Using the conversion, the average intracavity photon number $n_p$ reaches in excess of unity for emitters C, D and E, while $n_p$ maximizes below unity for the non-lasing emitters A and B at saturation.

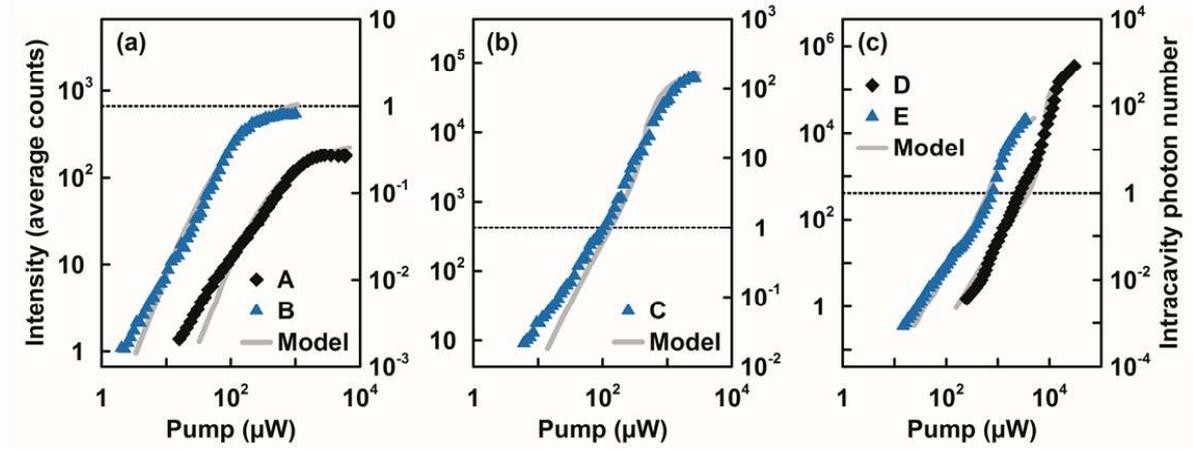

**Figure 3.** Input-output characteristics of QD-micropillar emitters for **(a)** LEDs, **(b)** cavity-enhanced LED and **(c)** lasers. The data points are from experiment and the grey curves are from theory. The left ordinate is average detector counts over integration period and the right ordinate is the calculated average intracavity photon number $n_p$ in the lasing mode. The dashed lines indicate the nominal lasing determined by $n_p = 1$.

For low excitation, all five emitters have very similar, slightly super-linear pump-power dependence. At higher excitations, the input-output dependence separates into strong saturation for the LEDs A and B, and nonlinear increase for the lasers D and E, as well as emitter C, the cavity-enhanced LED. The nonlinear increase, which signals the onset of stimulation emission, is considerably less pronounced than for conventional low-β lasers. This creates a need to look for further lasing confirmation in high-β situations.

Presently, the further confirmation often comes from the presence of a noticeable decrease in emission linewidth with increasing excitation. With conventional cavities where $\beta \ll 1$, the onset of linewidth narrowing occurs abruptly at lasing threshold, and typically scales inversely proportional to the excitation strength, consistent with the Schawlow-Townes description.[28] On the other hand, it is reported that with large β, the narrowing is less pronounced and then becomes undiscernible in the limiting case of $\beta = 1$.[20-25]

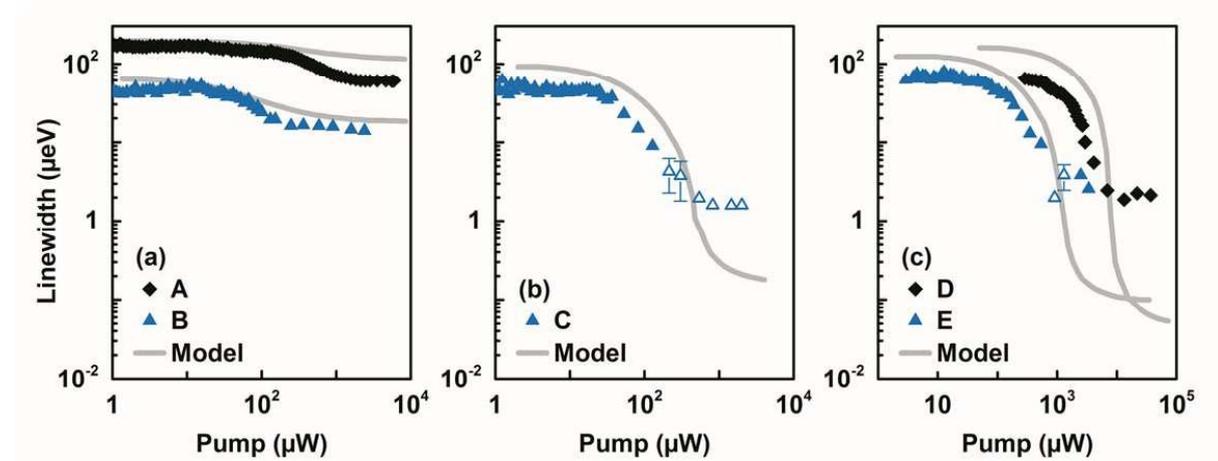

**Figure 4.** Spectral linewidth versus pump power of QD-micropillar emitters for (a) LEDs, (b) cavity-enhanced LED and (c) lasers. The grey curves are from theory and the data points are from experiment. Filled symbols indicate scanning Fabry-Pérot interferometer or spectrometer results, open symbols indicate results from analyzing $g^{(2)}(\tau)$ measurements.

Figure 4 shows plots of emission linewidth versus pump power for the five emitters. The data points are measured results obtained from a scanning Fabry-Pérot interferometer or the decay time of $g^{(2)}(\tau)$. First, the low pump-power plateaus show that emitters B, C, D and E have similar passive cavity linewidths of around 40-70 μeV, while emitter A has a larger passive cavity linewidth of roughly 140 μeV, indicating larger optical losses in this smallest diameter micropillar. With increasing pump power, emitters A and B show only slight decreases in linewidth, which support their assignment as LEDs. The linewidth reduction is partly from increased amplified spontaneous emission, which the model describes (grey curves in Fig. 4a) and partly from saturation of absorption losses, which the model does not take into account.

Figures 4b and 4c indicate appreciably greater linewidth narrowing with emitters C, D and E. The reduction of over two orders is due mostly to onset of stimulated emission, which is reproduced relatively well by the model. Discrepancy between theory and experiment arises at high excitation, with the theoretical curves indicating significantly narrower linewidths than observed experimentally. The measured linewidths are not spectrometer resolution limited, but may contain contributions from spectral fluctuations and temperature induced jitter at a timescale smaller than the sweep time (50ms per free spectral range) of the scanning Fabry-Pérot interferometer. Interestingly, the calculated minimum linewidth of approximately 0.1μeV, which translates to a coherence time of 13ns, matches well with the coherence time of ≈ 20 ns reported for a similar QD-micropillar laser.[39]

Other than the discrepancy caused by possibly the scanning Fabry-Pérot interferometer measurement sweep time, the CQED model is generally in agreement with experiment. The data points and curves in Fig. 3 show noticeable but acceptable differences. Varying $\gamma_{nl}$, $N_{QD}$ and $\eta$ to get better linewidth agreement would degrade the good agreement in input-output curves. We choose not to adjust the dephasing associated with $\langle b^\dagger b(\tau) \rangle$, which will change the linewidth independently of input-output behavior, but rather to base all effective scattering rates on reported results from experiments or quantum-kinetic calculations.[35,40,41] For the same reason, the polarization dephasing $\gamma$ is not treated as a free parameter in our curve fitting. Furthermore, the model neglects that for high-Q cavities, there may be excitation-dependent bleaching of absorption, which will result in a narrower linewidth.[42]

There remains the task of distinguishing between lasing and amplified spontaneous emission, with both phenomena exhibiting very similar excitation dependences of emission intensity and linewidth (compare Figs. 3b and 4b with Figs. 3b and 4c). A concern is that the plots in Figs. 2b and 3b may easily be mistaken as evidence for lasing in an ideal $\beta = 1$ device, because they show appreciable linewidth narrowing and a basically straight log-log input-output curve. To definitively separation of laser and cavity-enhanced LED operation, it is necessary to examine the equal-time intensity correlation $g^{(2)}(0)$. In recent years, there is considerable effort to improve techniques and equipment for determining $g^{(2)}(0)$ because of its important in characterizing single-photon and entangled-photon sources, and to a lesser extent, in verifying lasing, especially in the high-β regime.[9,34,42] Performing the necessary Hanbury Brown and Twiss (HBT) measurement remains labor and equipment demanding, for one faces the experimental issue of time resolution in the case of thermal light.[34,40] The challenge is that the coherence time of emission, which determines the timescales of $g^{(2)}(\tau)$ is usually much shorter than the timing resolution of the single-photon counting based HBT setups. This issue can be circumvented by using a streak-camera with dual time-base;[43] however, at the cost of a significantly lower quantum efficiency, which again sets constraints on measuring $g^{(2)}(\tau)$ in the thermal regime at low emission rates.

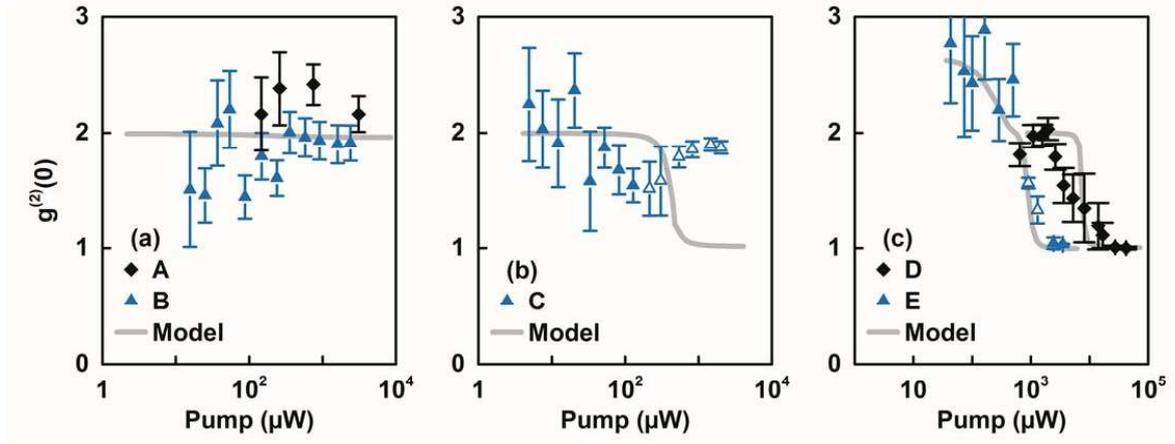

**Figure 5.** Equal-time intensity correlation versus pump power for QD-micropillar (a) LEDs, (b) cavity-enhanced LED and (c) lasers. The grey curves are from theory. The data points are from experiment, where the solid symbols indicate $g^{(2)}(\tau)$ analysis by integration and open symbols indicate fitting taking into account the instrument impulse response of the TCSPC setup.

For chaotic light, the Siegert relation links the normalized first-order correlation function $g^{(1)}(\tau)$ to the second-order photon autocorrelation function $g^2(\tau) = 1 + c|g^{(1)}(\tau)|^2$ with $c = 1$. A common approach is to model the intermediate regime between coherent and chaotic light with this function as well, where $0 < c < 1$. Homogenously broadened emission lines, as in our case, have a Lorentzian shape in the frequency spectrum; hence, the envelope of the normalized first order correlation function $g^{(1)}(\tau)$ is an exponential function decaying with the coherence time as time constant. Since the area under the $c|g^{(1)}(\tau)|^2$ function is conserved under convolution with the instrument impulse response function, it is now possible to estimate $c$, when the area and an estimate of the coherence time is given. Thus, by integrating over the measured raw $g^{(2)}(\tau)$ function, we estimate the original equal-time second-order photon autocorrelation function as $g^{(2)}(0) = 1 + \Delta E/(2\hbar) \int d\tau \left[g^{(2)}(\tau) - 1\right]$ for a Lorentzian spectrum with linewidth $\Delta E$ (FWHM). The data points in Fig. 5 shows the measured $g^{(2)}(0)$ versus pump power for all the devices. The data indicating thermal emission is obtained from analyzing the area under $g^{(2)}(\tau)$ function, while data showing the onset of stimulated emission or lasing are obtained directly by fitting the convolved model function to the HBT measurements. Emitter C, like the LEDs A and B, has $g^{(2)}(0)$ staying around 2 through the excitation range. On the other hand, micropillars D and E clearly exhibits a transition from thermal light ($g^{(2)}(0) \approx 2$) to coherent light ($g^{(2)}(0) \approx 1$). In all cases there is appreciable scattering in the data describing operation with thermal emission because of the time resolution and low emission rate challenges discussed in the previous paragraph. The grey curves are from the model, using the same input parameters are in Figs. 2 and 3.

Figure 5b depicts a peculiar behavior, where $g^{(2)}(0)$ for emitter C first reduces with increasing excitation, then turns around at roughly 100μW pump power and eventually returns to $g^{(2)}(0) = 2$ for pump power >1mW. Hence, in spite of the indications from Figs. 3b and 4b, emitter C does not represent a high-β laser, but is a good example for the importance of $g^{(2)}(0)$ measurements. We suspect the $g^{(2)}(0)$ behavior to be due to the carrier-density dependence of the dephasing rate, which has been reported to be strong in QD structures.[44,45] The difference between emitter C, and the lasers D and E may be that with a smaller diameter, it operates with higher carrier density in order to achieve gain. The present model neglects the carrier-density dependence of the dephasing coefficient, which explains the discrepancy between the calculated and measure results in Fig. 4b.

Additionally, micropillar E clearly shows $g^{(2)}(0) > 2$ before decreasing to unity with increasing excitation. The higher photon bunching may be from radiative emitter coupling that is present because of lower QD number.[46] Taking into account this contribution as discussed in the theory section, a relatively good fit to the experimental data was achieved, as can be seen in Fig. 5 (c). Radiative emitter coupling also affects the input-output curve for emitter E in Fig. 3(c) in the following manner. The transition from spontaneous emission exhibiting subradiance due to inter-QD coupling to stimulated emission destroying inter-QD correlations contributes to the jump in intensity at around $10^3$μW pump power, resulting in an extraction of β = 0.72 from fitting to experimental data. Neglecting the inter-QD correlations would give a much lower estimation of β = 0.23, which is of course is incorrect because of the unmistakable $g^{(2)}(0) > 2$ exhibited by emitter E as shown in Fig. 5(c). The appearance of subradiance depends sensitively on the number of QDs and their coupling to the common cavity mode.[46] This explains the different $g^{(2)}(0)$ power dependence of emitter D and E, which otherwise have very similar emission characteristics.

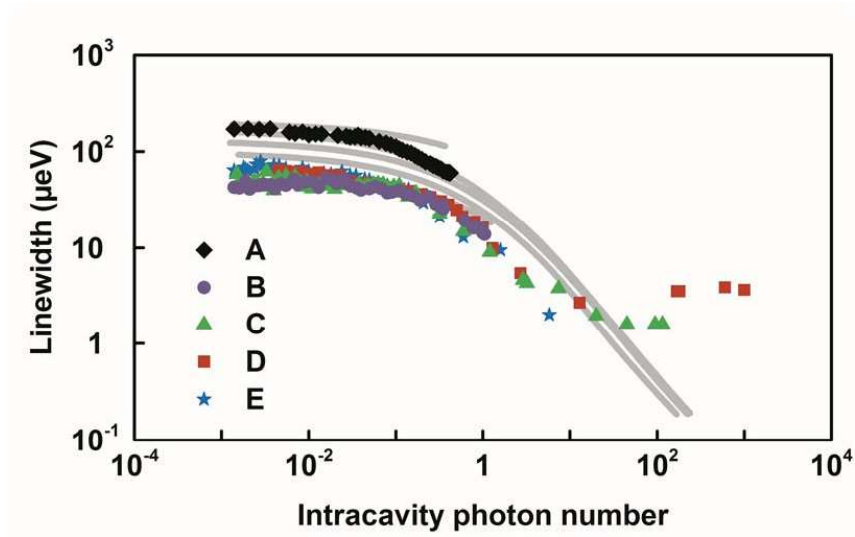

Figure 6. Spectral linewidth versus intracavity photon number in lasing mode. The data points are from experiment and the grey curves are from theory for (top to bottom) $Q = 8500, 25000$ and $32000$.

The measured and calculated information from the five QD-micropillars give a relatively complete picture of the broad range of emission properties exhibited by high-β devices. To understand the underlying physics governing the widely different behaviors, it is necessary to condense the information in Figs. 3 to 5. The first hint comes from noticing that the emission linewidth narrowing exhibited by all the emitters appears very similar when plotted versus intracavity photon number $n_p$, rather than the pump power. In Fig. 6, the curves calculated for $8500 < Q < 32000$ are essentially identical in terms of onset and slope of the line narrowing, when one factors out the different low excitation plateaus due to differences in passive cavity linewidth. The experimental data follows the same trend. In spite of the large number of experimental data points, the scatter is minimal.

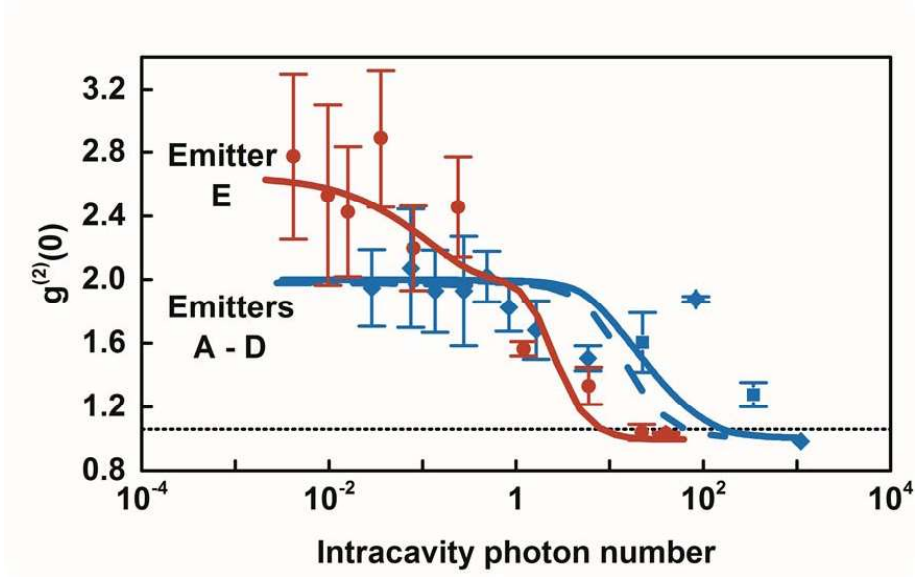

**Figure 7.** Equal-time intensity correlation versus intracavity photon number in lasing mode for all micropillar emitters. The data points are from experiment and the curves are from theory.

Extending upon Fig. 6, we plot $g^{(2)}(0)$ versus the intracavity photon number (Fig. 7). The blue curves are calculated using the broad range of the input parameters listed in Table 1 for emitters A to D ($8500 < Q < 32000, 15 < N_{QD} < 40, 0.03 < \eta < 0.3$ and $0.23 \leq \beta \leq 0.37$). That the curves are essentially overlapping provides strong indication of a single functional device-independent relationship regarding lasing. Within the standard deviation of the measurement, the experimental results support this claim. For clarity, the experimental data from the four micropillars is combined and presented in terms of an average and standard deviation at an averaged intracavity photon number. There is noticeably more scatter than in Fig 6 because of the challenges in measuring $g^{(2)}(0)$ for the low-intensity, thermal emission. Nevertheless, both theory and experiment appear to indicate a lasing threshold when the intracavity photon number reaches around $10^2$, independent of emitter configuration. The red curve and data points are for emitter E. Again they show an interdependence between photon correlations and photon number, in addition to a suggestion of a lower intracavity photon number necessary for reaching lasing threshold when superradiance is present.

Since linewidth narrowing will likely continue to be used as proof of lasing, it may be helpful to arrive at some quantitative indication of how much narrowing is necessary to support a lasing claim. Figure 7 shows the result from combining Figs. 5 and 6. The abscissa is the emission linewidth divided by passive cavity width. According to theory (grey curves), a greater than $10^2$ reduction is necessary. This refers to the fundamental linewidth determined solely by spontaneous emission. The measured data shows a less stringent requirement of roughly reduction to 16% of the passive cavity width, which is observed in emitter E. The apparent discrepancy arises because the measured linewidths of lasing emitters are further broadened by frequency jitter.

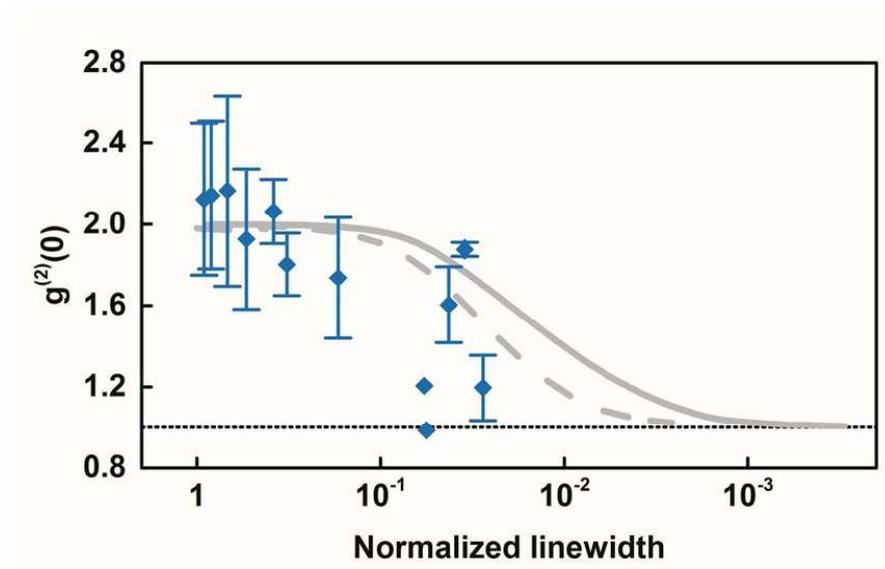

**Figure 8.** Equal-time intensity correlation versus linewidth reduction for all emitters. The linewidth reduction is defined as the emission linewidth divided by the passive cavity width. The data points are from experiment and the grey curves are from theory.

**CONCLUSIONS**

This paper describes research motivated by the considerable discussions involving near-unity spontaneous-emission factor ($\beta \lesssim 1$) emitters, in particular their unique emission properties during the transition from spontaneous emission to lasing. There are two aspects to our investigation. First is to learn, as much as possible, the emission properties of high-$\beta$ emitters. For the task, we fabricate, measure and model a wide variety of experimental configurations, ranging from LEDs to lasers. In addition to the usual characterization involving excitation dependences of emission intensity and linewidth, we measure and calculate the second-order intensity correlation $g^{(2)}(\tau)$ versus pump power. The importance to examine all three properties is illustrated very well by an emitter whose properties fall between those of an LED and a laser: a cavity-enhanced LED. Importantly, it can easily be mistaken for a laser because its intensity and linewidth excitation dependences are similar to those of lasers. However, its $g^{(2)}(0)$ clearly indicates thermal statistics. The importance of $g^{(2)}(0)$ is further reinforced by the results from a laser that exhibits a superradiant emitter coupling as it approaches the lasing threshold. Obviously, without knowing $g^{(2)}(0)$, the interesting property would have been missed, but equally important, an underestimation of the spontaneous emission factor would have resulted.

The second aspect of our research involves consolidating the vast amount of data so as to identify and understand the physics underlying the broad range of behaviors. Progress is made with the discovery that the experimental and theoretical results for all the emitters seemingly fit into two relationships: linewidth reduction versus intracavity photon number and $g^{(2)}(0)$ versus intracavity photon number. Hence, laser (or non-laser) action comes from achieving a given photon number above unity, and device parameters, such as Q and $\beta$ factors, quantum-dot density, dipole matrix element, carrier transport and optical-mode volume, determine the lasing threshold in that they affect the intracavity photon number.

**ACKNOWLEDGMENTS**

The research leading to these results has received funding from the European Research Council under the European Union's Seventh Framework ERC Grant Agreement No. 615613, from the German Research Foundation via the projects RE2974/5-1, Ka2318 7-1 and JA 619/10-3, and from the U.S. Department of Energy under Contract No. DE-AC04-94AL85000. WWC thanks the hospitality of the Technical University Berlin and travel support provided by the German Research Foundation via the collaborative research center 787.

**Figure captions**

**Figure 1. (Left)** Artist's impression of an optically pumped quantum-dot micropillar consisting of two GaAs/AlAs distributed Bragg reflectors (DBRs) and quantum dots (QDs) on a wetting layer. **(Right)** Scanning electron microscopy (SEM) image of 8 μm diameter micropillar, where part of the planarizing BCB layer (visible in the background) was mechanically removed to show the layers comprising the micropillar's DBRs.

**Figure 2.** Experimental setup. At the top left corner is a cryostat containing a sample that is pumped by a titanium:sapphire laser. The steady-state, constant wave (cw) pumped photoluminescence is analyzed with a grating spectrometer connected to a charge-coupled device (CCD), a scanning Fabry-Perot interferometer, a Hanbury Brown and Twiss (HBT) setup with single-photon counting modules (SPCMs) and a time-correlated single-photon counting (TCSPC) module. The cold mirror superimposes the pump laser onto the detection path, the pinhole provides spatial filtering, and the 850nm longpass filter blocks scattered pump laser light.

**Figure 3.** Input-output characteristics of QD-micropillar emitters for **(a)** LEDs, **(b)** cavity-enhanced LED and **(c)** lasers. The data points are from experiment and the grey curves are from theory. The left ordinate is average detector counts over integration period and the right ordinate is the calculated average intracavity photon number $n_p$ in the lasing mode. The dashed lines indicate the nominal lasing determined by $n_p = 1$.

**Figure 4.** Spectral linewidth versus pump power of QD-micropillar emitters for (a) LEDs, (b) cavity-enhanced LED and (c) lasers. The grey curves are from theory and the data points are from experiment. Filled symbols indicate scanning Fabry-Pérot interferometer or spectrometer results, open symbols indicate results from analyzing $g^{(2)}(\tau)$ measurements.

**Figure 5.** Equal-time intensity correlation versus pump power for QD-micropillar (a) LEDs, (b) cavity-enhanced LED and (c) lasers. The grey curves are from theory. The data points are from experiment, where the solid symbols indicate $g^{(2)}(\tau)$ analysis by integration and open symbols indicate fitting taking into account the instrument impulse response of the TCSPC setup.

Figure 6. Spectral linewidth versus intracavity photon number in lasing mode. The data points are from experiment and the grey curves are from theory for (top to bottom) $Q = 8500$, $25000$ and $32000$.

**Figure 7.** Equal-time intensity correlation versus intracavity photon number in lasing mode for all micropillar emitters. The data points are from experiment and the curves are from theory.

**Figure 8.** Equal-time intensity correlation versus linewidth reduction for all emitters. The linewidth reduction is defined as the emission linewidth divided by the passive cavity width. The data points are from experiment and the grey curves are from theory.

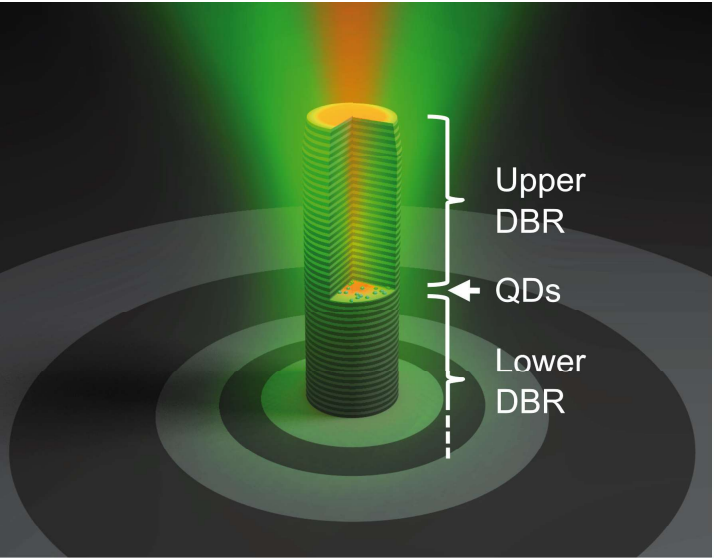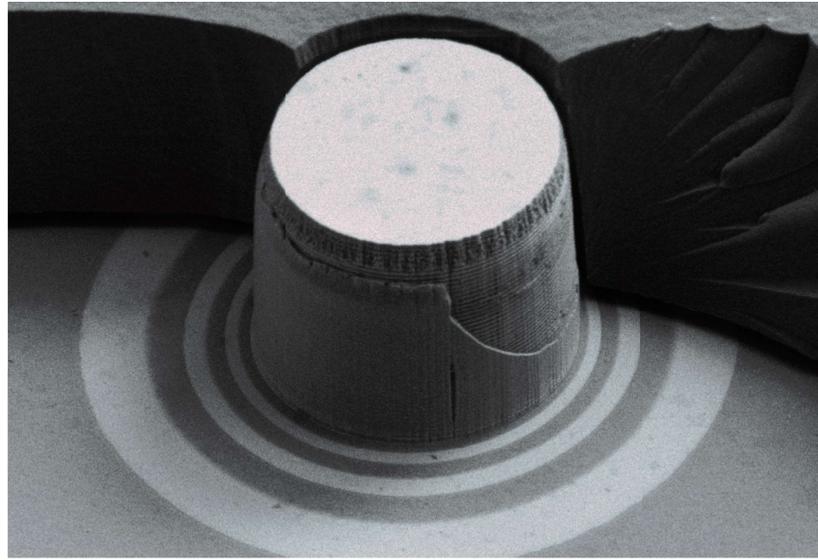

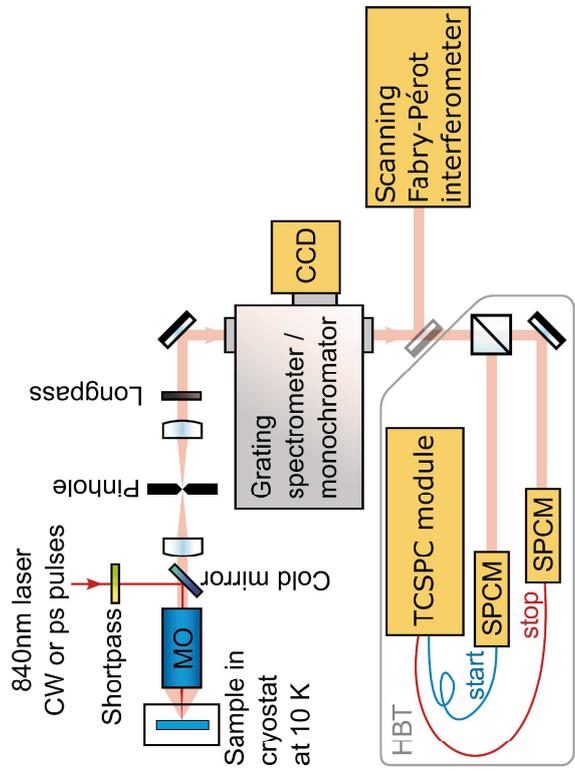

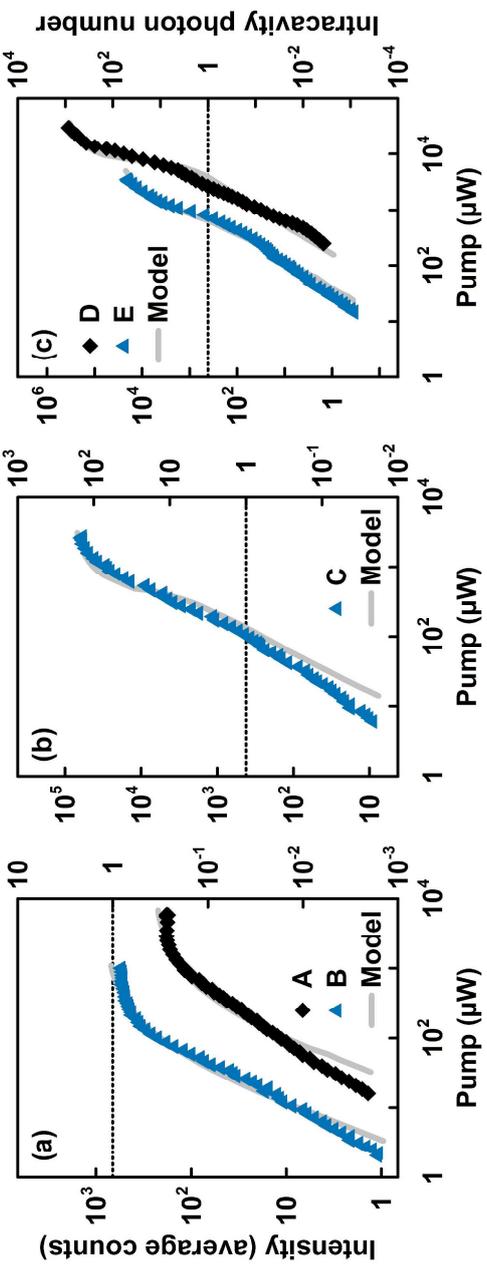

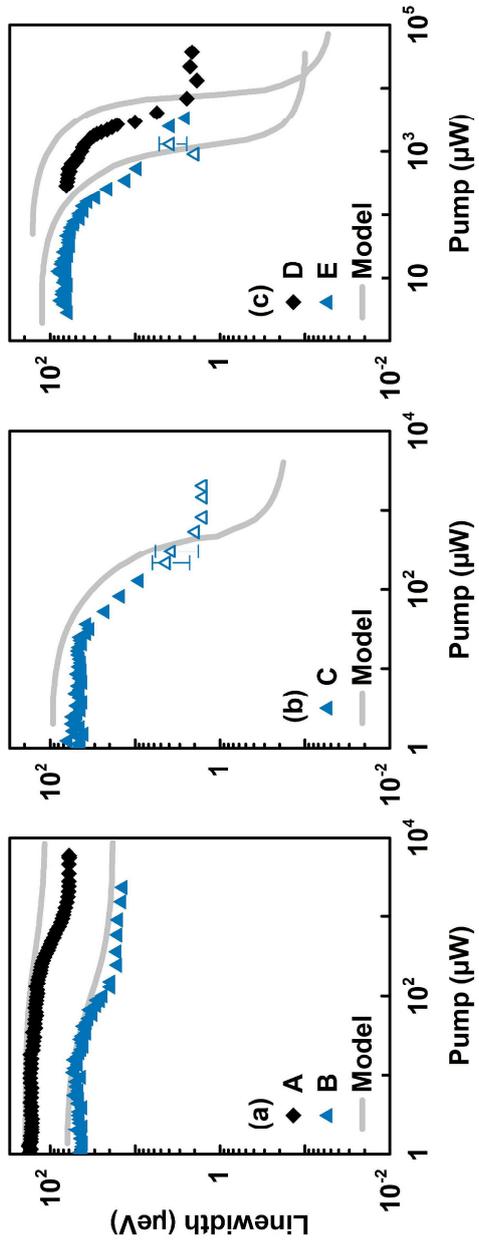

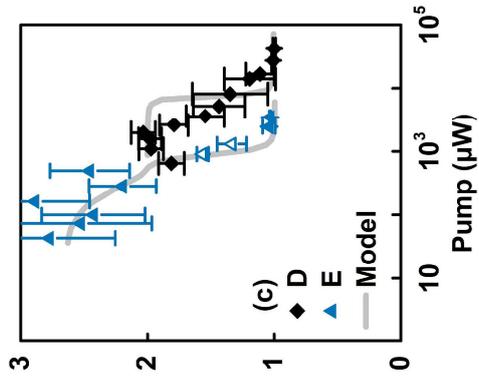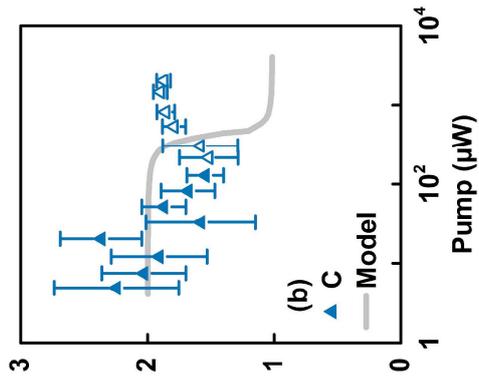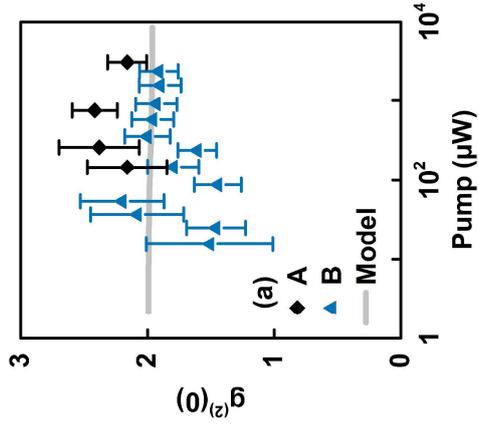

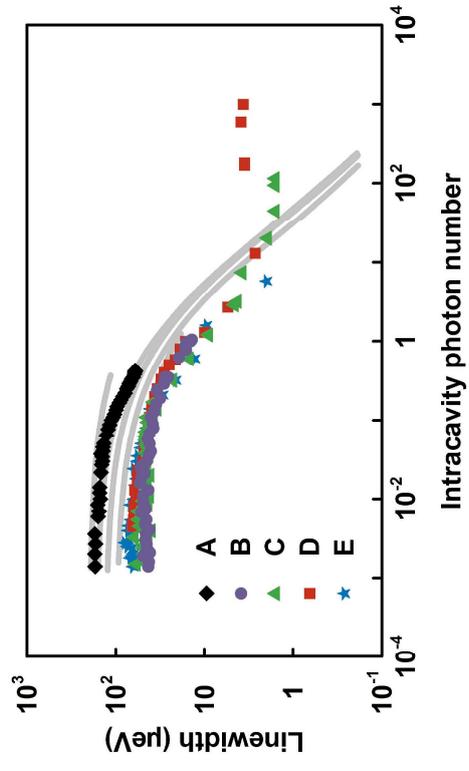

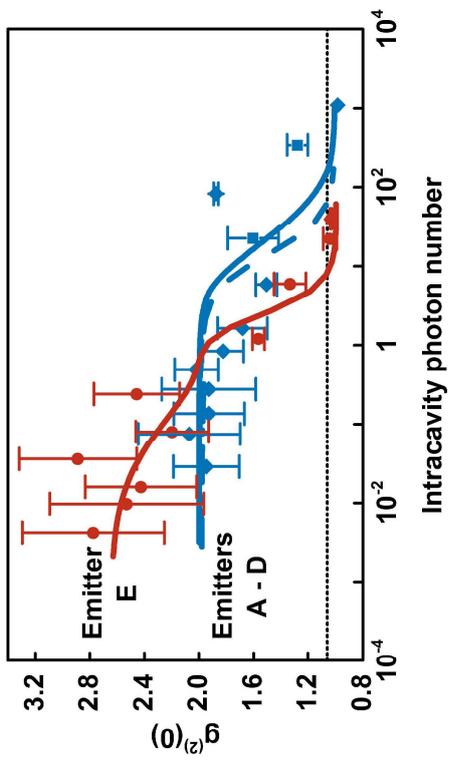

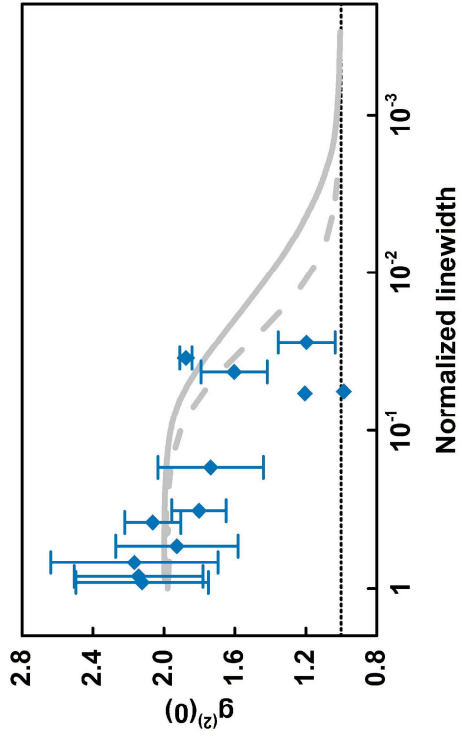